# Data synthesis improves detection of radiation sources in urban environments.

RR Flanagan[1], AG Osborne[1], MR Deinert[1,2]*
[1] Nuclear Science and Engineering, The Colorado School of Mines, Golden, CO 80401, USA
[2] Payne Institute for Public Policy, The Colorado School of Mines, Golden, CO, 80401, USA

**Abstract.** Radiation detectors in and around airports and other ports of entry play a crucial role in preventing nuclear terrorism. While these systems aim to detect and intercept nuclear materials before they enter a country, they are not foolproof. To address this, distributed sensors have been proposed to detect nuclear materials if they enter urban areas. Past work on small detector systems has shown that data fusion can improve detection. Here, we show how this could be done for a large detector network using Pearson's Method. We evaluate how a sensor network would perform in New York City using a combination of radiation transport and geographic information systems. We use OpenStreetMap data to construct a grid over the streets and analyze vehicle paths using pickup and drop off data from the New York State Department of Transportation. The results show that data synthesis in a large network (40k detectors) not only improves the time to the first detection but reduces the number of missed sources by up to 99%.

• Corresponding author: Mark Deinert, Department of Mechanical Engineering, The Colorado School of Mines: mdeinert@mines.edu

**Introduction.** The most significant danger associated with nuclear energy is the possibility of nuclear materials being used for malicious purposes, such as acts of terrorism [1]–[4]. The consequences of a nuclear weapon being detonated in a heavily populated area would result in devastating loss of life, and the use of a dirty bomb would cause widespread disruption [5]–[8]. To combat this, methods to detect nuclear materials in containers or luggage have been developed[9]–[11]. In the United States radiation detection systems for use around airports, ports, and border crossings have also been developed [12], [13], [14]. However, should these measures fail, it may still be possible for nuclear materials to be smuggled into the country. Once within the US, the materials could be transported to almost any location via highways and streets. Finding mobile nuclear materials in urban environments poses a more difficult problem than detecting them at border crossings [15], [16]. The problem is particularly acute for small or well shielded sources.

One possible solution to improve the detection of small and shielded sources is to enable a network of wirelessly connected, distributed sensors to operate in combination using data fusion. Brennan et al (2004) investigate this approach in general [17] by looking at a network of coupled acoustic, radar and radiation detectors to detect the movement of vehicles and demonstrated successful communication between all three types of detectors. King et al (2010) considered the distributed detection in the context of radiation detection with a network consisting of 13 mobile detectors was shown in [18] to enable a 1 mCi source to be detected with 90% likelihood in 10 minutes. In their study, 8 mobile detectors with fixed patrol routes, and 5 mobile detectors with random routes searched for a fixed source across a neighborhood. Bayesian as well as maximum likelihood estimation methods were used as data fusion techniques.

Data fusion, or machine learning, have been investigated for close range detection of radiation sources in suspected locations. Clustering, segmentation, and maximum likelihood estimator algorithms were studied by Liu and Sullivan (2016) [19] to detect multiple stationary sources in a 100 m-by-100 m area. This work showed a location detection accuracy of over 95%. In another study a Double Q-learning algorithm was used to navigate toward a radiation source, showing an improvement in location finding of 44% over a gradient search method [20]. This work again showed that data fusion techniques improve source detection, though with an alternative goal. A maximum likelihood optimization algorithm was used in [21] to find the location and strength of both moving and fixed sources in a 100 m-by-100 m square area. This method was shown to be able to pinpoint the location of the source down to 2m.

While past work has shown data fusion for distributed radiation detectors has significant potential for improving the detection of nuclear materials, it has been limited to small detector networks. Importantly, different sensors within a network can have different detection sensitivities, energy ranges, and response times. Distributed radiation detectors also generate large amounts of data, which can be difficult to store, transmit, and process in real-time. Because of this, approaches are needed which are computationally simple to implement and able to integrate data from different types of sensors. Here, we show that Pearson's method can be used effectively for synthesizing data to improve detection in a large network of mobile sources using a system of fixed and mobile detectors. We simulate up to 20,000 mobile detectors as in Flanagan et al (2021) [22] and 20,000 fixed detectors placed over the island of Manhattan, NY, with 0.01 and 0.001 Ci mobile sources. Patrol routes for mobile detectors are randomly generated to mimic the flow pattern of passenger vehicles such as taxis, Uber and Lyft cars. We show that data fusion improves the detection success rate and detection time compared to using individual sensors' measurements. We also find that placement of stationary detectors on the perimeter of Manhattan, and in physical locations with large view factors, significantly improves source detection. We also compare data fusion to non-network detectors assuming equal false positive rates.

**Methods.** A mobile radioactive source is simulated in conjunction with coupled stationary and mobile radiation detectors with the latter on simulated ride share vehicles in Manhattan, NY. Vehicles are assumed to transit public pedestrian streets only. Building geometries were obtained in shapefile format from OpenStreetMap [23]. Ride share data used for the mobile detectors is taken from pickup zones, drop off zones, and time stamps, made available by the New York Taxi and Limousine Commission. Random sampling of locations in the pickup and drop-offs zones is done to compute the pickup and drop-off locations and time stamps with the route-finding algorithm provided by the pyroute3 [24] library. Source routes start on the coast of Manhattan, bridges onto the island, or tunnels onto the island, with Madison Square Gardens as a destination. Source routes were also determined using pyroute3. The source and mobile detector routes were discretized into equally spaced time indexes with a $\Delta t$ of 2 s. The detector routes moved at a constant vehicle speed, which was computed using the total route length and duration. Geometric data were stored and manipulated using the Shapely [25] python library.

The simulated radioactive source was $Co^{60}$ with a strength of 0.01 Ci, emitting 1.17 and 1.33

MeV gamma rays. In the simulations, the source is shielded by 10 cm of lead. Additional shielding from the delivery vehicle itself was assumed to be 0.66 m of air and 1 cm of steel. The remaining distance between the source and the target is assumed to be air. For the purposes of this work buildings are opaque to radiation.

The gamma flux from the source at a given distance was approximated using a point source Greens function[26] which accounts for both the inverse square law and attenuation:

$$F(r) = \frac{Se^{-\sum_m^M r_m \mu_m}}{4\pi(\sum_m^M r_m)^2} * D_A * D_E \qquad (1)$$

Here, $F(r)$ {counts/s} is the strength of the source at distance $r$ {m}, $m$ is the material the radiation is passing through, $\mu_m$ {m$^{-1}$} is the linear attenuation coefficient for gamma rays in material $m$. $S$ {counts•s$^{-1}$} is the strength of the source after being attenuated by the shielding, and $r_m$ is the distance the radiation moves through material m, $D_E$ is the detector efficiency (0.2 or 20%) , and $D_A$ is the detector area (100cm$^2$). Sensors are assumed to be scintillating detectors that measure gross counts. Gross counts are also used for the background rate. The count rate leaving the shielding is calculated to be 431cps/cm$^2$, this number drops to 8.8cps/cm$^2$ at the edge of the vehicle.

Natural variations occur in both the background and source count rates. To account for this each timestep the background at each detector, and for each source, is computed by sampling a Poisson distribution based on their average count rates. An average background of 100 counts per second (cps) is used for Manhattan in this work, which is consistent with measured values [27]. Additionally, detectors ignore all count rates less than one standard deviation away from the mean, meaning only count rates over 110 cps are considered in this work. In reality the background count rate would change spatially and temporally within the city [28], [29], but that information is not currently available for Manhattan.

Two types of detector placement were simulated in this work, both using the same target in the interior of Manhattan and corresponding routes. In the first, 20,000 stationary detectors were placed randomly over the island of Manhattan. These detectors were combined with 20,000 mobile detectors. After being randomly placed, the stationary detectors were shifted to the closest area of high visibility determined by line of sight available to the detector. Most common areas of high visibility are the intersections between streets. The detector shifting is done by shifting each detector to the highest visibility area within 10m. The shifting process is shown in Figure 1.  In the second type of detector placement, the stationary detectors were instead placed along bridges entering the island or along roads near the coast (within 400 m). These detectors were also placed prioritizing high visibility locations. The same 20,000 mobile detectors were used in the second type of simulation. This formed a detector ring around Manhattan, through which mobile sources would have to pass.

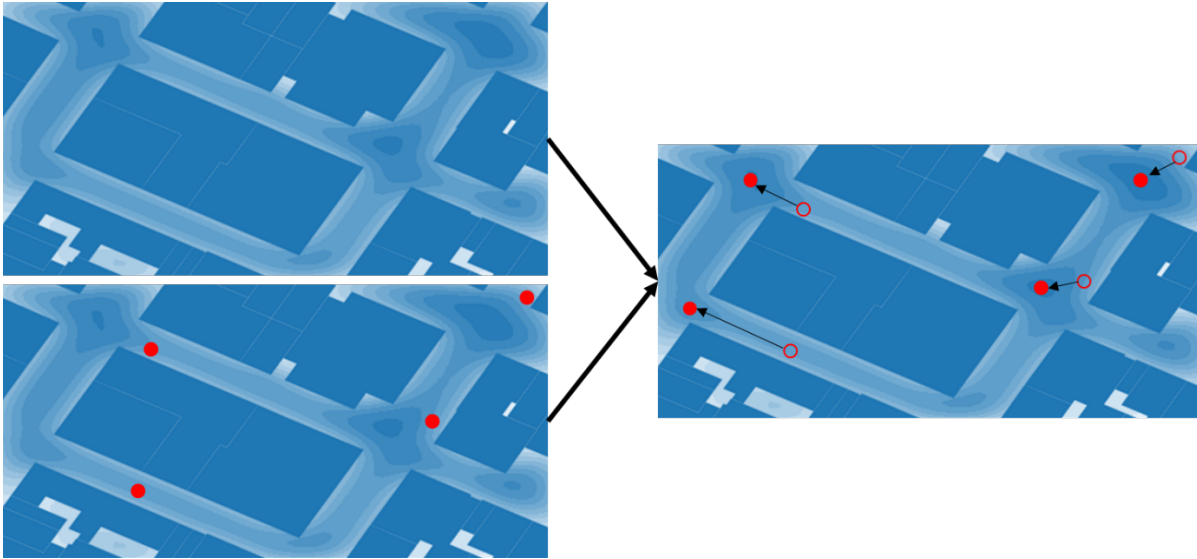

**Figure 1 The process for visibility shifting detectors to optimum locations.** The top left image represents a heat map of the amount of area each location can see. Darker blue colors indicate higher visibility. The bottom left image represents an example of randomly placed detectors (detectors cannot be placed in buildings). Detectors are represented by red circles. The image on the right shows how the detectors are shifted to the closest available area of high visibility within 10m of the original placement.

For each set of simulations, subsets of detectors (2.5k – 20k) were randomly sampled from the 20,000 mobile and 20,000 fixed detectors. Each subset was sampled evenly between the stationary detectors and mobile detectors. The sampling was performed 10,000 times for all subsets in both simulation styles (randomly placed vs strategically placed). For example, 5k detectors were randomly selected and the time to detection is calculated for each route. This process was repeated 10,000 times and the average time to detection for each route is calculated. Duplicated samples were dropped and replaced to ensure no identical detector subsets existed in the 10,000 random samples. Figure 2 gives the flowchart for the algorithm.

One thousand source routes from the coast of Manhattan to the target were generated. Routes that started on locations that did not have a nearby road had their starting location shifted to the nearest point on a road. Vehicles were assumed to be traveling at 10 mph [30] as a baseline, with random full stop moments (0 mph) to simulate traffic and traffic lights. The locations of the vehicles were recorded every 2 seconds.

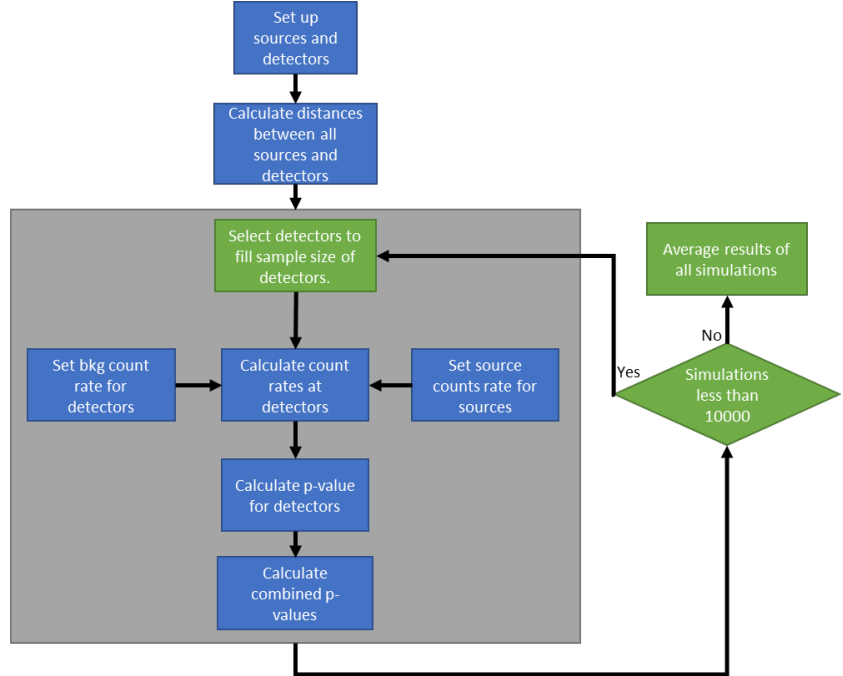

**Figure 2**. **Method for determining average number of detections and detection time for each source route**. A representation of the sampling method used for each source route. Selected detectors are split equally among mobile and stationary detectors.

The time until the source was detected, and the number of missed sources was used as a comparison between the methods used in this study. The three methods used were a non-network system, a network using Pearson's method [31] to combine probabilities, and a network using Stouffer's method [32] to combine probabilities. Both network methods use only detections that occur in succession. This was done to prevent detectors within a localized area simultaneously registering a background variation at the same time and considering it a radioactive source. P-values for detectors were calculated based on a Poisson distribution of the background count rate (100 cps). The two types of detector placement are compared against each other. This comparison aims to show the benefits of strategically placed stationary detectors vs randomly placed stationary detectors.

*Non-Network Detection.* As a baseline for comparison each detector was treated as an individual entity with no knowledge of the other detectors. In this system if the detector registered a count rate of 3 standard deviations above average background of Manhattan it registered a detection of radioactive material. Individual detectors did not store information from previous timesteps and only alerted if the current time step registered a count rate of 3 standard deviations above average background.

*Pearson's Method.* Pearson's method is a data fusion technique that combines p-values into a single Chi-squared statistic that can be used to calculate a new p-value for the combined statistical tests. If the p-value of this test was less than 0.01 the network registered a detection of a radioactive source. Pearson's method chi square score was calculated using:

$$\chi^2_{2I} \sim -2 \sum_{i=1}^{I} \log(1 - p_i) \qquad (2).$$

Here, χ represents the chi-squared score of the new distribution, *I* is the total number of tests, and $p_i$ is the p-value of the i[th] test[31].

*Stouffer's Method.* Stouffer's method is another data fusion technique to combine p-values over a series of statistical tests. The basis of the method is a Z-score, setting it apart from the Pearson's method. If the p-value of this test was less than 0.01 the network registered a detection of a radioactive source. Stouffer's method z score is calculated using the following:

$$z = \sum_{i=1}^{I} \frac{z_i}{sqrt(I)} \quad (3)$$

where z represents the z-score of the new distribution, *I* is the total number of tested z-scores, and $z_i$ is the z-score of the i[th] test [32].

*Source Detection in Network Methods.* Starting with time step 0 all detectors that register above the threshold count rate were isolated. Consider detector $d_i$. All detectors within range (twice the possible distance covered by a source) of $d_i$ in the next time step were checked for count rates above threshold. A combined p-value was calculated for $d_i$ and detectors in time step t+1 ($d_{i\_n}$). This chain was continued until there were no above threshold detections in the succeeding timesteps. If a chain of detectors registered a combined p-value less than 0.01, the location and time of the final detector in the chain was stored. The location and time were tested against the source route to determine if a source was within range of the detection. If so, the source was considered detected, if not, a false positive was registered. This method ensures that detectors do not know the location of a source until after a possible detection occurs. This process is demonstrated in Figure 3.

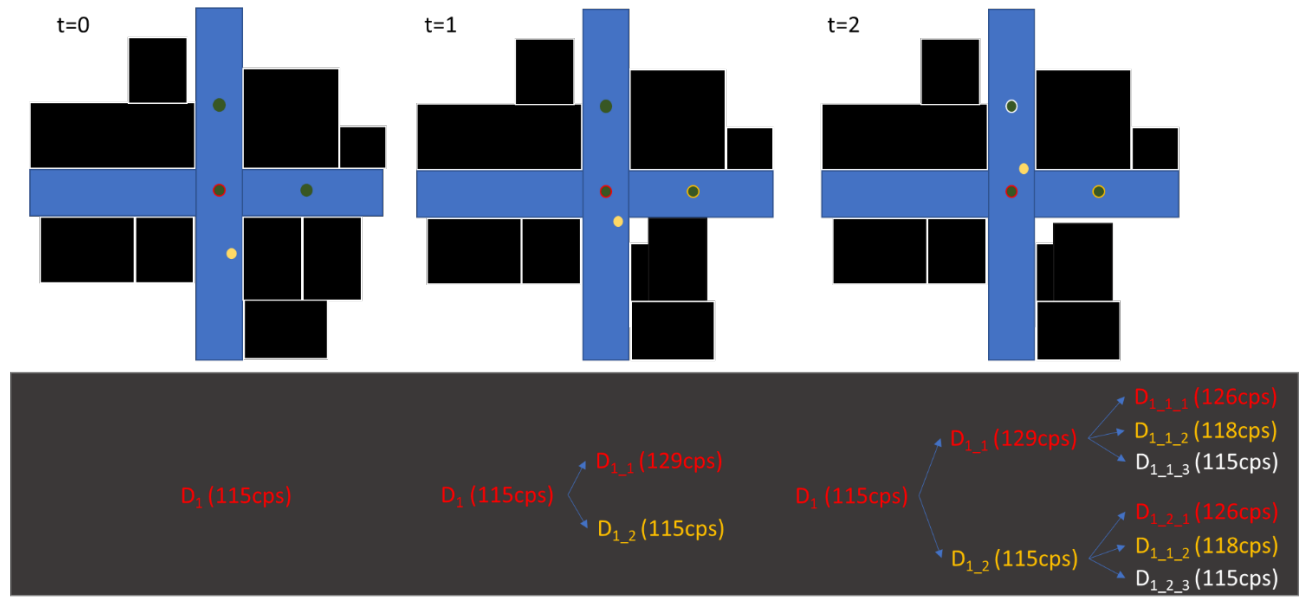

**Figure 3**. **Method for determining detector chains**. Detector chains develop as a tree of detections over the cutoff count rate. Here detectors are colored green with color outlines representing detections above count rate. The source vehicle is shown in yellow. The tree of detections is used to generate the chains for calculating combined p-value.

**Results.** Table 1 shows the results for the average detection time and the average number of missed source routes for randomly placed stationary detectors. Here half the detectors are mobile and half are fixed. The results show that the Pearson's method produces the greater reduction in detection time of the two network methods. Increasing the number of detectors also reduces the average detection time across all methods. Detectors are split evenly between mobile detectors and stationary detectors. Additionally, Pearson's method also has the lowest number of missed source routes on average missing ~20 fewer than Stouffer's method and ~60 lower than the non-network method. This work does not investigate a combination of the two methods, that is an area for future investigation. For this work an average false positive rate of 1.84% was observed.

When considering the number of missed routes, 5k detectors with Pearson's method performed as well as 12.5k non-networked detectors. For Stouffer's method 5k networked detectors performed as well as 7.5k non-networked detectors. The comparisons are approximately the same for average detection time.

**Table 1. The average time to detection and average misses for the three detector methods using randomly placed stationary detectors.**

| | Average Detection Time (s) | | | Average Misses | | |
|---|---|---|---|---|---|---|
| Number of Detectors (half mobile, half fixed) | Pearson's | Stouffer's | No Network | Pearson's | Stouffer's | No Network |
| 5k | 213.19 | 219.26 | 234.41 | 380.55 | 398.94 | 440.36 |
| 7.5k | 197.15 | 203.94 | 220.93 | 342.79 | 363.01 | 403.27 |
| 10k | 190.38 | 197.32 | 215.16 | 327.60 | 348.33 | 387.78 |
| 12.5k | 187.33 | 194.26 | 212.50 | 321.32 | 342.19 | 380.82 |
| 15k | 185.95 | 192.85 | 211.34 | 318.47 | 339.40 | 377.75 |
| 17.5k | 185.30 | 192.20 | 210.76 | 317.12 | 338.11 | 376.28 |
| 20k | 184.99 | 191.88 | 210.49 | 316.55 | 337.53 | 375.59 |

The effects of the reduction can be seen in Figure 4, which shows the differences between the non-network method to the statistical methods for a network of 20,000 mobile and 20,000 fixed detectors. Figure 4 assumes the randomly placed detector simulation. This figure represents an example route, not all routes see the same reduction in detection time. The example source route in Figure 4 demonstrates the effectiveness of the statistical method and highlights the time improvement. Notice that while both the statistical methods, and the non-network method, detect the source on the way to its target, the statistical methods do so much earlier. This demonstrates that even though all three methods detect the source, there is still benefit to the statistical methods due to the improved detection time.

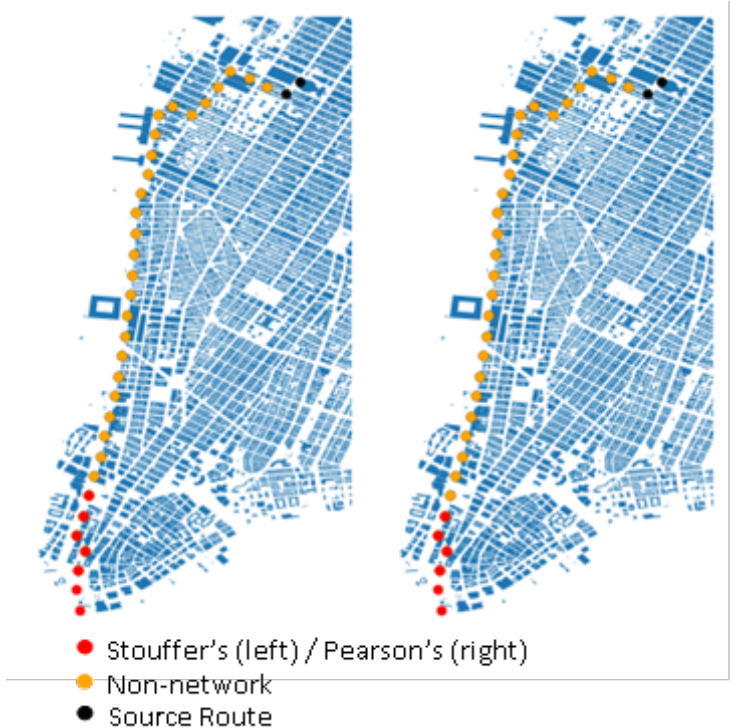

**Figure 4. Point of detection for the different methods used on the same route.** Each dot represents a point on the source trip every 5$^{th}$ timestep (10 s). Here, 40,000 randomly placed detectors (20,000 mobile and 20,000 fixed). The red dots represent how far the source got before being detected by a network method. The yellow dots end where the source was detected using non-networked detectors, and finally the black dots represent the full source route.

Table 2 shows the results for the strategically placed detector simulations. These results show improvement in all categories compared to the randomly placed detector simulations.

**Table 2. The average time to detection and average misses for the three detector methods using strategically placed stationary detectors.**

| Number of Detectors (half mobile, half fixed) | Average Detection Time (s) | | | Average Misses | | |
|---|---|---|---|---|---|---|
| | Pearson's | Stouffer's | No Network | Pearson's | Stouffer's | No Network |
| 5k | 150.65 | 168.64 | 185.51 | 124.31 | 158.12 | 284.62 |
| 7.5k | 116.25 | 130.29 | 176.23 | 55.32 | 76.37 | 264.69 |
| 10k | 97.66 | 109.12 | 174.47 | 26.17 | 39.70 | 248.81 |
| 12.5k | 86.43 | 96.18 | 167.49 | 12.87 | 21.67 | 236.37 |
| 15k | 78.88 | 87.38 | 164.14 | 6.77 | 12.23 | 231.64 |
| 17.5k | 73.58 | 81.14 | 160.86 | 3.54 | 7.03 | 213.11 |
| 20k | 69.67 | 76.49 | 154.42 | 1.96 | 4.38 | 210.98 |

Comparing the results in Table 1 and in Table 2 the benefit of strategically placing the detectors is immediately obvious. Moving the detectors to be more clustered around locations on the entry points of the island, and along the coastal roads, means that sources that are detected are done so much sooner. This also reduces the number of missed routes. This is because increasing the number of detectors a source route passes only increases the chance that a route is close enough to a detector to be detected. Importantly, the impact of data fusion increases dramatically with the strategic placement of fixed detectors. The data synthesis methods benefit from the increased number of detectors a source experiences because additional p-values improve the results of those methods.

The results in Table 2 show that data synthesis using Pearson's method makes 5k networked detectors as effective as 20k non-networked ones regarding detection time. When considering the frame of average number of misses, 5k networked detectors significantly outperformed 20k non-network detectors. Pearson's method provided a reduction of 41.1% and Stouffer's method a reduction of 25.1%. The difference between networked and non-networked detectors grows with their number. For 20k detectors, networking them with Pearson's method reduced misses by more than 99%.

**Table 3. The average time to detection and average misses for the three detector methods using strategically placed stationary detectors with equal false positive rates.**

|  | Average Detection Time (s) | | Average Misses | |
| --- | --- | --- | --- | --- |
| Number of Detectors (half mobile, half fixed) | Pearson's | No Network | Pearson's | No Network |
| 5k | 150.65 | 176.54 | 124.31 | 234.08 |
| 7.5k | 116.25 | 172.23 | 55.32 | 174.75 |
| 10k | 97.66 | 167.36 | 26.17 | 168.13 |
| 12.5k | 86.43 | 150.91 | 12.87 | 137.66 |
| 15k | 78.88 | 138.0 | 6.77 | 112.11 |
| 17.5k | 73.58 | 128.54 | 3.54 | 94.03 |
| 20k | 69.67 | 121.12 | 1.96 | 80.38 |

Table 3 is a comparison between the Pearson's Method and no network method, but the false positive rate for Pearson's method is used to set the p-value threshold, or false positive rate for the no network method. The results of this table show an improvement in the detection time and rate between the methods. However, the network method still outperforms the non-network method at an equivalent false positive rate.

An important result to highlight from this work is the number of detectors required to detect a source moving through Manhattan. Under the best conditions with 5k detectors, 12.4% of source routes are missed. As the number of detectors decreases to 1k, the number of missed routes is 49.3%. Finally, while the results presented above are for a 10mCi source, the detection rate at 5k detectors was repeated at 1mCi. Using the lower source strength, the detection rate for Pearson's method drops to 43.6% and for the non-network method the detection rate drops to 18.3%. For 20k detectors the detection rate for Pearson's method drops to 63.4% and for the non-network method the detection rate drops to 31.7%.

There are several limitations with this study. An average background of 100 counts per second is used for Manhattan in this work, which is consistent with measured values [27]. However, it is known from other work that the background rate within cities varies with location [28], [29], and this is likely true for Manhattan as well, though they were not available for this work. Mobile detectors could help to map the background radiation in a city [33] and future work along these lines could be used to improve results [34]–[36]. A full radiation transport model for buildings was not included here. Instead, buildings are assumed to be completely opaque and this reduces source visibility (a conservative estimate). Correspondingly, full radiation transport models for the attenuation of cars from different angles are not included in the current work. Instead, the shielding of the source is assumed to dominate attenuation. Source routes in this work are driven along optimum travel time routes toward their destination. However, it is possible that an attacker would instead drive in an unoptimized method to avoid locations where detectors are likely. Additionally, points of entry onto the island that bypass the coast (e.g. heliports) are not considered. This work also does not consider the financial cost of detectors and the methods investigated work require at least 15k detectors to have less than a 1% false negative rate. However, it is possible that other data fusion methods, perhaps using spectra, would reduce this number.

**Conclusions.** The probability of using nuclear materials for an attack in a given city is likely unknowable, but the risk is ever present. Ensuring that the most likely targets of these weapons, cities, are prepared is important to preventing such an attack. Mobile and stationary sources that work in isolation can detect these sources using count rate alone. However, for the type of detection investigated here (based on count rate alone) the results demonstrate that the performance of detection is increased when data synthesis is done. This is particularly true when fixed detectors are placed in such a way as to ensure a source would have to pass multiple ones. This improvement holds true even in the case when the networked and non-networked detectors have equal false positive rates. This work also demonstrates that increasing the number of detectors available increases detection. However, for the type of data fusion used here, smartly placing stationary detectors provides a much larger benefit, with just 5k strategically placed detectors performing better than 20k randomly placed detectors.